# Černý type automata and rank conjecture

**Igor Rystsov**

National technical university of Ukraine (NTUU), Ukraine

E-mail address: haryst49@gmail.com

**Abstract.** The aim of this paper is to prove the Černý conjecture and the rank conjecture for Černý type automata and monoids. A transformation monoid is said to be Černý type if it is generated by a simple idempotent and a regular group of permutations. We prove Černý conjecture for the Černý type synchronizing automata and the rank conjecture for the Černý type transformation monoids. In particular, we obtain the tight bound for the reset threshold of Černý type synchronizing monoids.

## 1. Introduction

A finite automaton can be regarded as a transformation monoid with a prescribed set of generators. A finite automaton is synchronizing if its monoid contains a constant map, whose action sends all states to one state. The reset threshold of a synchronizing automaton is the minimum length of its constant maps. Synchronizing automata are most famous due to the Černý conjecture, according to which a synchronizing n-state automaton has a constant map of length at most $(n-1)^2$ [1].

Here, we define the reset threshold of a finite transformation monoid, which does not depend from a particular set of generators. We generalize Černý conjecture to the rank conjecture for transformation monoids and prove it for Černý type monoids.

A similar approach was proposed in [2], where was considered some varieties of monoids and, in fact, was proved the rank conjecture for their rank thresholds. The theory of transformation monoids was developed in [3].

## 2. Preliminaries

Let $S$ be a finite set consisting of $n$ states, where $n > 2$. The full transformation monoid $\text{End}(S) = \{f \mid f: S \to S\}$ on a set $S$ is the monoid whose elements are all the mappings from $S$ to itself and whose operation is composition.

The elements of the full monoid will be called maps for simplicity. A transformation monoid $M$ on $S$ is a submonoid of $\mathrm{End}(S)$.

The symmetric group $\mathrm{Sym}(S) = \{g \mid g: S \leftrightarrow S\}$ on a set $S$ is the group whose elements are all the permutations of $S$ and whose operation is composition. A permutation group $G$ on $S$ is a subgroup of $\mathrm{Sym}(S)$. The degree of $G$ is the cardinality of $S$. Note that $\mathrm{Sym}(S)$ is the subgroup of all invertible maps in $\mathrm{End}(S)$. The difference $\mathrm{End}(S) \setminus \mathrm{Sym}(S)$ consists of the singular maps on $S$.

We write maps on the right of their argument, so that $(s)f$ is the image of $s$ under the map $f$. This avoids the reordering of maps under its composition. The image of $f$ is the subset of states $\mathrm{im}(f) = \{(s)f : s \in S\}$. Its complement $\mathrm{cim}(f) = S \setminus \mathrm{im}(f)$ is the co-image of $f$. The image of a subset $T \subseteq S$ is the subset $(T)f = \{(s)f : s \in T\}$. The rank $\mathrm{rk}(f)$ of $f$ is the cardinality of its image $\mathrm{rk}(f) = |\mathrm{im}(f)|$. The number $\mathrm{crk}(f) = |\mathrm{cim}(f)| = n - \mathrm{rk}(f)$ is the co-rank of $f$.

Maps of co-rank zero are permutations in the symmetric group $\mathrm{Sym}(S)$. Maps of nonzero co-rank are the singular maps in $\mathrm{End}(S) \setminus \mathrm{Sym}(S)$. Note that each singular map merges at least one pair of states. A simple singular $f$ is a map of the unit co-rank $\mathrm{crk}(f) = 1$. A simple singular merges exactly one pair of states. A constant map $f$ is a map of the unit rank $\mathrm{rk}(f) = 1$, which merges all states.

Let $M$ be a monoid, then a homomorphism $\varphi: M \rightarrow \mathrm{End}(S)$ of monoids defines an action of $M$ on $S$ by the formula for all $s \in S$ and $x \in M$:
$$(s)x = (s)\varphi_x, \qquad (1)$$
where $\varphi_x = (x)\varphi$. This action satisfies the conditions for all $s \in S$, and $x, y \in M$:
$$(s)(x \cdot y) = ((s)x)y, \qquad (s)1_M = s, \qquad (2)$$
where $1_M$ is the identity of $M$ [3, 4]. The action is faithful if the homomorphism $\varphi$ is injective. Particularly, if $M \leq \mathrm{End}(S)$, then we have a faithful action $i_M: M \hookrightarrow \mathrm{End}(S)$, since the identity embedding $i_M$ is a monomorphism. For our purposes, it is sufficient to consider only faithful (strict) actions, but a more general point of view sometimes will be very useful. The advantage of this approach is that the same monoid may acts on several different sets.

The image $(s,t)f$ of a pair $(s,t) \in S \times S$ under the map $f \in \text{End}(S)$ is defined by components $(s,t)f = ((s)f, (t)f)$. The image of a relation $\rho \subseteq (S \times S)$ is the subset of pairs $(\rho)f = \{(s,t)f : (s,t) \in \rho\}$. A relation $\rho$ is invariant for a transformation monoid $M$, if $(\rho)f \subseteq \rho$ for each $f \in M$. The complement of a relation $\rho$ is denoted by $\overline{\rho} = (S \times S)\backslash\rho$.

Let $\varepsilon \subseteq (S \times S)$ be an equivalence relation and $S/\varepsilon = \{S_1, \cdots, S_r\}$ be the corresponding equivalence classes (partition on $S$), then the number of classes $r$ is its rank $\text{rank}(\varepsilon) = |S/\varepsilon| = r$. A transversal (or section) of a partition $S/\varepsilon$ is a subset $T \subseteq S$ which meets every class $S_i$ in a single point $|S_i \cap T| = 1, 1 \leq i \leq r$. We say that $\varepsilon$ is t-invariant for a transformation monoid $M$, if $(T)f$ is a transversal of $S/\varepsilon$ for each transversal $T$ of $S/\varepsilon$ and $f \in M$ [4].

**Lemma 2.1.** An equivalence $\varepsilon \subseteq (S \times S)$ is t-invariant for a transformation monoid $M$ if and only if the complement $\overline{\varepsilon}$ is invariant for $M$.

**Proof.** If $\text{rank}(\varepsilon) = 1$, then the lemma is trivially true. Let $T = \{t_1, \cdots, t_r\}$ be a transversal for $\varepsilon$, where $r = \text{rank}(\varepsilon) > 1$. Assume that $(T)f$ is not a transversal of $S/\varepsilon$ for some $f \in M$, then by the box principle (Dirichlet principle) there is a class $S_i \in S/\varepsilon$, which contain images of two states $(t_j)f$ and $(t_k)f$, where $(t_j, t_k) \in \overline{\varepsilon}$. Therefore, the complement $\overline{\varepsilon}$ is not invariant for $M$. The converse statement is obvious, since any pair of states $(t_1, t_2) \in \overline{\varepsilon}$ can be extended to a transversal. Thus, the lemma is proved. $\square$

An equivalence $\varepsilon \subseteq (S \times S)$ is a congruence of a transformation monoid $M$ if it is invariant for $M$. If $G$ is a permutation group on $S$ and $\varepsilon$ its congruence, then we have $(\varepsilon)g = \varepsilon$ for all $g \in G$, since $|(\varepsilon)g| = |\varepsilon|$. From this follows that $(\overline{\varepsilon})g = \overline{\varepsilon}$ for all $g \in G$. Thus, we obtain the corollary from Lemma 2.1.

**Corollary 2.2.** Each congruence of a permutation group is t-invariant.

The rank $\text{rk}(M)$ of a transformation monoid $M \leq \text{End}(S)$ is the minimal rank of its maps $\text{rk}(M) = \min_f \{\text{rk}(f) : f \in M\}$. It is well-known from a semigroup folklore [3] that the minimal ideal $I(M)$ of a transformation monoid $M$ consists of the maps with minimal rank $I(M) = \{f : \text{rk}(f) = \text{rk}(M)\}$. A transformation monoid $M$ called synchronizing if $\mathrm{rk}(M) = 1$. In this case, its minimal ideal consists of constant maps.

**Lemma 2.3.** If an equivalence $\varepsilon \subseteq (S \times S)$ is t-invariant for a transformation monoid $M$, then $\mathrm{rk}(M) \geq \mathrm{rank}(\varepsilon)$.

**Proof.** Let $T$ be a transversal of the partition $S/\varepsilon$ and $f \in M$, then $(T)f \subseteq \mathrm{im}(f)$. From this follows that $\mathrm{rank}(\varepsilon) = |(T)f| \leq \mathrm{rk}(f)$ for each $f \in M$. Thus, $\mathrm{rank}(\varepsilon) \leq \mathrm{rk}(M)$ and the lemma is proved. □

Let $M \leq \mathrm{End}(S)$ be a transformation monoid and $T \subseteq S$ be a subset of states. Suppose that $(T)f \subseteq T$ for each $f \in M$, then we have for each $t \in T$:

$$(t)f = (t)f|_T, \qquad (3)$$

where $f|_T$ is the restriction of a map $f \in M$ on $T$. From Eq. (2), (3) we have the following equalities for each $t \in T$ and $f, g \in M_A$:

$$(t)(f \cdot g)|_T = (t)(f \cdot g) = ((t)f)g = ((t)f|_T)g|_T = (t)(f|_T \cdot g|_T). \qquad (4)$$

Let us put $M|_T = \{f|_T : f \in M\}$, then from Eq. (4) follows that the mapping $\varphi: f \to f|_T$ will be a homomorphism of monoids $\varphi: M \to M|_T$. Thus, the monoid $M$ acts on the subset $T$ according Eq. (1), (3). The monoid $M|_T$ is called a transformation submonoid of $M$ on the subset $T$.

Let $\varepsilon \subseteq (S \times S)$ be a congruence of a transformation monoid $M \leq \mathrm{End}(S)$ and $S/\varepsilon = \{S_1, \cdots, S_r\}$ be the corresponding congruence classes. Let us fix $S_i \in S/\varepsilon$ and $f \in M$, then from inclusion $(\varepsilon)f \subseteq \varepsilon$ follows that there is the unique class $S_j$, which satisfy the condition $(S_i)f \subseteq S_j$. Hence, we may define a map $f|_\varepsilon$ on $S/\varepsilon$ by the formula:

$$(S_i)f|_\varepsilon = S_j, \quad \text{if } (S_i)f \subseteq S_j, \ 1 \leq i \leq r. \qquad (5)$$

If $(S_i)f|_\varepsilon = S_j$ and $(S_j)g|_\varepsilon = S_k$, then from (2) and (5) it is easy to see that:

$$(S_i)(f \cdot g)|_\varepsilon = S_k = (S_i)(f|_\varepsilon \cdot g|_\varepsilon).$$

So, the mapping $\varphi: f \to f|_\varepsilon$ will be the homomorphism of monoids $\varphi: M \to M|_\varepsilon$, where $M|_\varepsilon = \{f|_\varepsilon : f \in X\}$. The monoid $M|_\varepsilon$ is called a transformation factor-monoid of $M$ by the congruence $\varepsilon$.

## 3. Automata

Further, we use some notions, which was introduced in [5]. Denote by $\langle X \rangle$ the transformation monoid on $S$, which is generated by the maps $X \subseteq \text{End}(S)$.

**Definition 3.1.** An automaton $A$ is a pair $A = (S, X)$, where $S$ is a set of states and $X$ is a subset of maps $X \subseteq \text{End}(S)$.

The transformation monoid $M_A = \langle X \rangle$ is called the transition monoid of an automaton $A$. The rank of an automaton $A = (S, X)$ by definition equals to the rank of its transition monoid $\text{rk}(A) = \text{rk}(M_A)$. Therefore, an automaton $A$ is synchronizing if its transition monoid $M_A$ is synchronizing.

Now let us fix a set of generators $X$ in a transformation monoid $M$. This allows to define the measure of complexity for a map factorization on generators. The length $l_X(f)$ of a map $f \in M$ by respect to $X$ be the smallest length of a representation the map $f$ as a composition (product) of maps from $X$. The introduced metrics has the following property for all $f, g \in M$:

$$l_X(f \cdot g) \leq l_X(f) + l_X(g) \ . \tag{6}$$

The reset threshold $\text{rt}(A)$ of a synchronizing automaton $A$ defined as the minimum length of its constants $\text{rt}(A) = \min_c \{ l_X(c) : c \in I(M_A) \}$. The well-known Černý conjecture [1] formulated in the following way.

**Conjecture 3.1.** The reset threshold of a synchronizing $n$-state automaton $A = (S, X)$ satisfies the inequality $\text{rt}(A) \leq (n-1)^2$.

This conjecture can be generalized as follows. We may consider an automaton $A = (S, X)$ as a transformation monoid $M = M_A$ with a fixed set of generators $X$. Then we define the rank threshold $\text{rt}_X(M)$ of an automaton $(M, X)$ as the minimum length of its maps with the minimal rank:

$$\text{rt}_X(M) = \min_f \{ l_X(f) : f \in I(M) \} \ .$$

The rank threshold of a transformation monoid $M \leq \text{End}(S)$ defined as the maximum rank threshold of an automaton $(M, X)$ over all sets of generator $X \subseteq M$:

$$\text{rt}(M) = \max_{X \subseteq M} \{ \text{rt}_X(M) : \langle X \rangle = M \} \ . \tag{7}$$

Then the rank conjecture can be formulated in the following way [2].

**Conjecture 3.2** (rank conjecture). For each $n$-state transformation monoid $M$ of the rank $r$ holds the inequality $\mathrm{rt}(M) \leq (n-r)^2$.

## 4. Černý type automata and monoids

A map $f$ called an idempotent if $f \circ f = f$. An idempotent $f$ acts identically on its image $(s)f = s$ for each $s \in \mathrm{im}(f)$. A simple idempotent $f$ is an idempotent, which is also a simple singular $\mathrm{crk}(f) = 1$. In this case, there is exactly one state $e_f \in \mathrm{cim}(f)$ (excluded state), which transforms into a different state $d_f$ (duplicate state) under the map $f$. All other states transfer into themselves. Further, we denote by $f$ a simple idempotent.

A transformation monoid $M$ on a set $S$ is transitive if for every pair of states $s, t \in S$ there is a map $f \in M$ such that $(s)f = t$. A permutation group $G$ on a set $S$ is transitive if it is a transitive transformation monoid. The subset of permutations $G_s = \{g \in G : (s)g = s\}$ is called the stabilizer of the state $s$ in $G$. A permutation group $G$ on a set $S$ is semiregular if for each $s \in S$ its stabilizer equals to the identity permutation $G_s = \{1_S\}$ [6].

A permutation group $G$ on a set $S$ is regular if it is semiregular and transitive. In this case $|G| = |S| = n$ and for every two states $s, t \in S$ there is the unique permutation $g \in G$ such that $(s)g = t$. In other words, the following condition holds for each $s \in S$ and $g, h \in G$:

$$(s)g = (s)h \rightarrow g = h. \tag{8}$$

If $Y$ is a set of generators for a regular permutation group $G \leq \mathrm{Aut}(S)$, then for any $g \in G$ the following inequality holds:

$$l_Y(g) \leq |G| - 1 = n - 1. \tag{9}$$

The index of a subgroup $H \leq G$ in a group $G$ is denoted by $|G:H|$.

**Definition 4.1.** A transformation monoid $M \leq \mathrm{End}(S)$ is called Černý type if $M = \langle \{f\} \cup G \rangle$, where $f$ is a simple idempotent and $G$ is a regular permutaion group.

Further, we denote by $y$ the unique permutation of a Černý type monoid $M$, which satisfies the condition:

$$(e_f)y = d_f, \quad y \in G. \tag{10}$$

Let us put $H = \langle y \rangle$ and $r = |G:H|$, then the right cosets $H_i = Hg_i$, $1 \leq i \leq r$, of the group $G$ by the subgroup $H$ form a partition $\{H_1, \cdots, H_r\}$ on the group $G$. Then from Formula (8) follows that the following subsets form a partition on a set $S$:

$$S_i = (e_f)H_i, 1 \leq i \leq r, \tag{11}$$

**Lemma 4.1** Let $M = \langle \{f\} \cup G \rangle$ be a Černý type monoid and $\varepsilon$ be the equivalence with classes $S_i$, $1 \leq i \leq r$, which are defined by Eq. (11), then $\varepsilon$ is a t-invariant congruence of $M$.

**Proof.** Let us put $S_1 = (e_f)H$, then $d_f, e_f \in S_1$, and we have properties:

$$(S_1)f = S_1 \setminus \{e_f\}, \quad (S_i)f = S_i, 2 \leq i \leq r. \tag{12}$$

Further, for every $g \in G$ the partition $\{(S_1)g, \cdots, (S_r)g\}$ will be the permutation of classes $S_i$, since $(S_i)g = (e_f)H_i g$, $1 \leq i \leq r$. Hence, $\varepsilon$ is a congruence of $G$ and from Eq. (12) we conclude that $\varepsilon$ is also a congruence of $M$.

Now let $(s_j, s_k) \in \bar{\varepsilon}$, where $s_j \in S_j$, $s_k \in S_k$, then from Eq. (12) we deduce that $(s_j)f \in S_j$, $(s_k)f \in S_k$, so $((s_j)f, (s_k)f) \in \bar{\varepsilon}$. Hence, $(\bar{\varepsilon})f \subseteq \bar{\varepsilon}$. Further, since $\varepsilon$ is a congruence of $G$ we have $(\bar{\varepsilon})g = \bar{\varepsilon}$ for each $g \in G$. Thus, $\bar{\varepsilon}$ is invariant for $M$ and according Lemma 2.1 $\varepsilon$ will be t-invariant for $M$. The lemma is proved. $\square$

From Lemma 4.1 and Lemma 2.3 we obtain the following statement.

**Corollary 4.2.** Let $M = \langle \{f\} \cup G \rangle$ be a Černý type monoid and $y \in G$ satisfies Eq. (10), then $\mathrm{rk}(M) \geq |G:\langle y \rangle|$.

In the sequel, we need some properties of images and co-images of maps. From Eq. (2) we have the following property for all maps $, w \in \mathrm{End}(S)$:

$$\mathrm{im}(v \cdot w) = (\mathrm{im}(v))w. \tag{13}$$

**Lemma 4.3.** Let $f$ be a simple idempotent, $g \in \mathrm{Sym}(S)$, and $w \in \mathrm{End}(S)$, then $\mathrm{cim}(w \cdot g) = (\mathrm{cim}(w))g$ and the property holds:

$$d_f \in \mathrm{im}(w) \rightarrow \mathrm{cim}(wf) = \mathrm{cim}(w) \cup \{e_f\}. \tag{14}$$

**Proof.** We have by definition that $\{\mathrm{im}(w), \mathrm{cim}(w)\}$ is a partition on $S$, then $\{(\mathrm{im}(w))g, (\mathrm{cim}(w))g\}$ is also a partition on $S$. From this and Eq. (13) we obtain:

$$(\text{cim}(w))g = S\backslash(\text{im}(w))g = S\backslash\text{im}(wg) = \text{cim}(wg).$$

Now let $d_f \in \text{im}(w)$, then we have the following alternative:

$$\text{im}(wf) = \begin{cases} \text{im}(w)\backslash\{e_f\}, & \text{if } e_f \in \text{im}(w) \\ \text{im}(w), & \text{if } e_f \in \text{cim}(w) \end{cases}.$$

In the first case $\text{im}(wf) = \text{im}(w)\backslash\{e_f\}$ and taking the complement from both sides of this equality, we obtain $\text{cim}(wf) = \text{cim}(w) \cup \{e_f\}$. In the second case, we deduce that $\text{cim}(wf) = \text{cim}(w) = \text{cim}(w) \cup \{e_f\}$. Thus, the lemma is proved. $\square$

An automaton $A = (S, X)$ is Černý type if its monoid is Černý type $M_A = \langle X \rangle = \langle \{f\} \cup G \rangle$.

**Theorem 4.4.** Let $A = (S, \{f\} \cup Y)$ be a Černý type automaton with $n$ states and the regular cyclic group $G = \langle Y \rangle = \langle y \rangle$, where $y$ satisfies Eq. (10), then $\text{rk}(A) = 1$ and $\text{rt}(A) \leq (n-1)^2$.

**Proof.** We have $G = \langle y \rangle = \langle h \rangle$ and $e_f = (d_f)h$, where $h = y^{-1} = y^{n-1}$. Thus, taking into account Formula (8), we obtain the equality:

$$S\backslash\{d_f\} = \{(d_f)h, \cdots, (d_f)h^{n-1}\}. \qquad (15)$$

Further, by definition we have $\text{cim}(f) = \{e_f\} = \{(d_f)h\}$. From this and Lemma 4.3 follows that $\text{cim}(fh) = (\text{cim}(f))h = \{(d_f)h^2\}$. Note that $d_f \neq (d_f)h^2$, since $n > 2$, so $d_f \in \text{im}(fh)$. Hence, from Formula (14) we have:

$$\text{cim}(fhf) = \text{cim}(fh) \cup \{e_f\} = \{(d_f)h, (d_f)h^2\}.$$

Continuing to argue in this way from Eq. (15), we obtain by induction on $k$, $1 \leq k \leq n-2$, the equality:

$$\text{cim}(f(hf)^k) = \{(d_f)h, \cdots, (d_f)h^{k+1}\}. \qquad (16)$$

Let us put $k = n-2$ and $w = f(hf)^{n-2} = (fh)^{n-2}f$, then from Eq. (15), (16) we conclude that $\text{cim}(w) = S\backslash\{d_f\}$ and the following equality holds:

$$\text{rk}(A) = \text{rk}((fh)^{n-2}f) = 1. \qquad (17)$$

Let $X = \{f\} \cup Y$, then from Properties (6), (9), we have the upper bound $l_X(fh) \leq l_X(f) + l_Y(h) \leq 1 + (n-1) = n$. From this and Eq. (17) we obtain the estimation for the reset threshold of $A$:

$$\text{rt}(A) \le l_X(fh)(n-2) + l_X(f) \le n(n-2) + 1 = (n-1)^2.$$

Thus, the theorem is proved. □

This theorem follows from many earlier papers [4, 7, 8, 9], but we will need it later. From Corollary 4.2 and Theorem 4.4 we obtain the following statement.

**Theorem 4.5.** A Černý type monoid $M = \langle\{f\} \cup G\rangle$ is synchronizing if and only if $G = \langle y \rangle$, where $y$ satisfies the Eq. (10).

**Proof.** If $\text{rk}(M) = 1$, then by Corollary 4.2 we have:

$$1 = \text{rk}(M) \ge |G:\langle y\rangle| \ge 1.$$

From this, we conclude that $|G:\langle y\rangle| = 1$ and $G = \langle y\rangle$. Hence, $M = \langle\{f\} \cup G\rangle = \langle\{f, y\}\rangle$. The converse statement follows from Theorem 4.4. □

Let $M = \langle\{f, y\}\rangle$ be a Černý type synchronizing monoid, where $(e_f)y = d_f$. Without loss of generality, we may assume, that $S = \{0, 1, \cdots, n-1\}$, $f = (0, 1, \cdots, n-2, 0)$ is the simple idempotent and $y = (1, 2, \cdots, n-1, 0)$ is the cyclic permutation, as in Černý automata [10]. So, $d_f = 0$ and $e_f = n-1$.

Consider the map $fyf$ and note that it is a simple singular $(S)fyf = (S\setminus\{n-1\})yf = (S\setminus\{0\})f = S\setminus\{n-1\} = \text{im}(f)$. Consider also the cyclic permutation $v = (1, 2, \cdots, n-2, 0, n-1)$, where the last state $n-1$ is fixed. Therefore, $v^{n-1} = 1_S$, where $1_S$ is the identity map on $S$.

**Lemma 4.6.** The following equalities hold $(fyf)^{m(n-1)} = f$ and $(fyf)^{m(n-1)+k} = (fyf)^k$, for all $m \ge 1$, $1 \le k \le (n-2)$.

**Proof.** At first, we show that $fyf = fv$. Indeed, it is easy to verify that the following equalities are true:

$$(s)fyf = (s)fv = (s+1) \bmod (n-1), \quad 0 \le s \le n-1.$$

From this, by induction on $k$ we obtain for each $k \ge 1$:

$$(s)(fyf)^k = (s)fv^k = (s+k) \bmod (n-1), \quad 0 \le s \le n-1. \quad (18)$$

Further, we have $(fyf)^{m(n-1)} = (fv^{n-1})^m = (f \cdot 1_S)^m = f^m = f$ for all $m \ge 1$. From this, we obtain $(fyf)^{m(n-1)+k} = f \cdot (fyf)^k = (fyf)^k$, for all $m \ge 1$ and $1 \le k \le (n-2)$. Thus, the lemma is proved. □

The following theorem complements Theorem 4.4.

**Theorem 4.7.** Let $A = (S, \{x\} \cup Y)$ be an automaton with $n$ states, the simple singular $x = (fyf)^k$ and the regular cyclic group $G = \langle Y \rangle = \langle y \rangle$, where $(e_f)y = d_f$, then $A$ is a Černý type synchronizing automaton and $\operatorname{rt}(A) \leq (n-1)^2$.

**Proof.** We have by definition $x \in \langle \{f, y\} \rangle$ then $M_A \subseteq \langle \{f, y\} \rangle$, since $\langle Y \rangle = \langle y \rangle$. On the other hand, from Lemma 4.6 we have $x^{n-1} = (fyf)^{k(n-1)} = f$. Hence, we obtain $M_A = \langle \{x\} \cup \langle Y \rangle \rangle = \langle \{f\} \cup \langle y \rangle \rangle = \langle \{f, y\} \rangle$. Thus, from Theorem 4.5 we conclude that $A$ is a Černý type synchronizing automaton.

We may assume that $1 \leq k \leq (n-2)$ according Lemma 4.6, since if $k = m(n-1)$, where $m \geq 1$, then $(fyf)^k = f$ and this case was considered in Theorem 4.4. Further, we use the natural order on the set $S = \{0, 1, \cdots, n-1\}$. From Eq. (18) we see that on the sequence of states $(n-1-k, \cdots, n-1, 0, \cdots n-2-k)$ the map $x$ increases from $0$ to $n-2$. Moreover, the map $x$ is strictly monotonic on this interval except for one place, where $(n-1)x = (0)x = k$.

At first, we apply the map $x$ to $A$. If $k < n-2$, we perform the first stage and transfer the subset of states $\{0, \cdots, m\}$, where $k+1 \leq m \leq n-2$, by the permutation $y^{n-1-k}$ to the sequence $(n-1-k, \cdots, n-1, 0, \cdots, m-k-1)$ and then apply $x$:

$$(\{0, \cdots, m\}) y^{n-1-k} x = \{0, \cdots, m-1\}. \tag{19}$$

We repeat this operation $(n-2-k)$ times and as the result, we obtain the map $w_1 = x(y^{n-1-k}x)^{n-2-k}$, for which $\operatorname{im}(w_1) = \{0, \cdots, k\}$ on the strength of Eq. (19). If $k = n-2$, then $w_1 = x$ and we immediately proceed to the second stage.

At the second stage, we transfer the subset of states $\{m, \cdots, k\}$, $0 \leq m < k$, by the permutation $y^{n-k}$ to the sequence of states $(n-k+m, \cdots, n-1, 0)$ and then apply the map $x$:

$$(\{m, \cdots, k\}) y^{n-k} x = \{m+1, \cdots, k\}.$$

We repeat this operation $k$ times and as the result, we obtain the map $w = w_1 w_2$:

$$w = w_1 w_2 = x(y^{n-1-k}x)^{n-2-k}(y^{n-k}x)^k,$$

which is a constant map $\operatorname{im}(w) = \{k\}$. From this, we have the following equalities:

$$\operatorname{rk}(A) = \operatorname{rk}(w_1 w_2) = \operatorname{rk}(x(y^{n-1-k}x)^{n-2-k}(y^{n-k}x)^k) = 1. \tag{20}$$

Let us put $X = \{x\} \cup Y$, then from Formula (6) and Eq. (20) we have:

$$\mathrm{rt}(A) \leq l_X(w) \leq l_X(y^{n-1-k}x)(n-2-k) + l_X(y^{n-k}x) \cdot k + 1 \,. \tag{21}$$

Further, from Ineq. (6), (9) we deduce:

$$l_X(y^{n-1-k}x) \leq l_Y(y^{n-1-k}) + l_X(x) \leq (n-1) + 1 = n \,,$$

and analogously $l_X(y^{n-k}x) \leq n$. From this and Ineq. (21) we obtain the bound for the reset threshold $\mathrm{rt}(A) \leq n(n-2) + 1 = (n-1)^2$ and the theorem is proved. $\square$

Now we consider a set of generators in a Černý type monoid.

**Lemma 4.8.** Let $X$ be a set of generators of a Černý type monoid $M = \langle \{f\} \cup G \rangle$ then there is a simple singular $z \in X$ such that $z = g_0 f g_1$ or $z = g_0(fyf)^k g_1$, where $g_0, g_1 \in G$ and $y \in G$ satisfies Eq. (10).

**Proof.** We divide the set $X$ into two parts $X = Y \cup Z$, where $Y$ are permutations, which generates the group $G = \langle Y \rangle$, and $Z$ are singulars. Consider a factorization $f = gzw$ by generators from $X$, where $g \in G$, $z$ is the first singular from $Z$ in this factorization and $w \in \langle X \rangle$. Then we have:

$$\mathrm{rk}(z) = \mathrm{rk}(f) = n - 1. \tag{22}$$

Let us also fix a shortest factorization of $z$ by generators from $\{f\} \cup G$:

$$z = g_0 f g_1 f \cdots g_k f g_{k+1} \,, \tag{23}$$

where $g_i \in G \setminus \{1_S\}$, $0 \leq i \leq k+1$.

If $k = 0$, then $z = g_0 f g_1$, and the lemma is proved in this case. Further, we assume that $k \geq 1$ in Eq. (23). Consider the map $fgf$, where $g \in G \setminus \{1_S\}$. Then from Lemma 4.3 follows that $\mathrm{cim}(fg) = \{(e_f)g\}$. If $g \neq y$, then $(e_f)g \neq d_f$ and $d_f \in \mathrm{im}(fg)$. Hence, from Lemma 4.3 we obtain the property:

$$\mathrm{cim}(fgf) = \{e_f, (e_f)g\}, \quad \mathrm{rk}(fgf) = n - 2, \text{ for each } g \in G \setminus \{1_S, y\} \,.$$

From this and Eq. (22), we conclude that $z = g_0(fyf)^k g_{k+1}$, and the lemma is proved $\square$

From Theorem 4.4 follows that a Černý type synchronizing automaton $A = (S, \{f, y\})$ satisfies Conjecture 4.1. Now we consider the reset threshold of a Černý type synchronizing monoid.

**Theorem 4.9.** The reset threshold of a Černý type synchronizing monoid $M = \langle\{f, y\}\rangle$, where $n = |\langle y \rangle|$, satisfies the inequality $\text{rt}(M) \leq (n-1)^2$.

**Proof.** Let $X$ be a set of generators of a monoid $M = \langle\{f, y\}\rangle$, for which $\text{rt}_X(M) = \text{rt}(M)$. Recall that the permutation $y$ satisfies Eq. (10) and $n = |\langle y \rangle| = |S|$. We divide $X$ into two parts $X = Y \cup Z$, where $Y$ are permutations, which generates the cyclic group $G = \langle Y \rangle = \langle y \rangle$, and $Z$ are singulars. From Lemma 4.8 follows that there is a simple singular $z \in Z$ such that $z = g_0 f g_1$ or $z = g_0 (fyf)^k g_1$, where $g_0, g_1 \in G$.

Let us put $h = y^{n-1}$. Suppose that $z = g_0 f g_1$, then we have $z g_1^{-1} h g_0^{-1} = g_0 f g_1 g_1^{-1} h g_0^{-1} = g_0 f h g_0^{-1}$. From this, we obtain by induction that $(z g_1^{-1} h g_0^{-1})^{n-2} = g_0 (fh)^{n-2} g_0^{-1}$. Hence, we have the map $w = (z g_1^{-1} h g_0^{-1})^{n-2} z = g_0 (fh)^{n-2} f g_1 = g_0 u_1 g_1$. So, from Eq. (17) we deduce:
$$|(S)w| = |(S) g_0 (fh)^{n-2} f_1 g_1| = |(S)(fh)^{n-2} f g_1| = \text{rk}((fh)^{n-2} f) = 1.$$
Further, from Ineq. (6), (9) we have the bound:
$$l_X(z g_1^{-1} h g_0^{-1}) \leq l_X(z) + l_Y(g_1^{-1} h g_0^{-1}) \leq 1 + (n-1) = n.$$
From this, we obtain in this case the following estimation:
$$\text{rt}_X(M) \leq l_X(w) = l_X((z g_1^{-1} h g_0^{-1})^{n-2} z) \leq n(n-2) + 1 = (n-1)^2. \quad (24)$$

Now consider the case when $z = g_0 x g_1$, where $x = (fyf)^k$. We may assume that $1 \leq k \leq (n-2)$ according Lemma 4.6. Further, we have $z g_1^{-1} y^{n-1-k} g_0^{-1} = g_0 x g_1 g_1^{-1} y^{n-1-k} g_0^{-1} = g_0 x y^{n-1-k} g_0^{-1}$. From this, we obtain by induction that:
$$(z g_1^{-1} y^{n-1-k} g_0^{-1})^{n-2-k} = g_0 (x y^{n-1-k})^{n-2-k} g_0^{-1},$$
and analogously $(z g_1^{-1} y^{n-k} g_0^{-1})^k = g_0 (x y^{n-k})^k g_0^{-1}$. Thus, we have the following equalities:
$$w = (z g_1^{-1} y^{n-1-k} g_0^{-1})^{n-2-k} (z g_1^{-1} y^{n-k} g_0^{-1})^k z =$$
$$g_0 (x y^{n-1-k})^{n-2-k} (x y^{n-k})^k x g_1 = g_0 x (y^{n-1-k} x)^{n-2-k} (y^{n-k} x)^k g_1 = g_0 u g_1,$$
where $u = x (y^{n-1-k} x)^{n-2-k} (y^{n-k} x)^k$. So from Eq. (20) we deduce:
$$|(S)w| = |(S) g_0 u g_1| = |(S) u g_1| = \text{rk}(u) = 1.$$
Similar to the first case, we have the following bound:
$$l_X(z g_1^{-1} y^{n-1-k} g_0^{-1}) \leq l_X(z) + l_Y(g_1^{-1} y^{n-1-k} g_0^{-1}) \leq 1 + (n-1) = n,$$

and analogously $l_X(zg_1^{-1}y^{n-k}g_0^{-1}) \leq n$. Hence, we obtain in this case the following estimation:

$$\mathrm{rt}_X(M) \leq l_X(w) = l_X((zg_1^{-1}y^{n-1-k}g_0^{-1})^{n-2-k}(zg_1^{-1}y^{n-k}g_0^{-1})^k z) \leq$$
$$\leq n(n-2-k) + n \cdot k + 1 = n(n-2) + 1 = (n-1)^2 \,.$$

From this and Formula (24) we conclude that $\mathrm{rt}(M) = \mathrm{rt}_X(M) \leq (n-1)^2$, and the theorem is proved. □

Let us consider the Černý function (or Shannon function) $C_1(n)$ for the class of Černý type synchronization monoids, which defined by formula:

$$C_1(n) = \max_M \{\mathrm{rt}(M) : M = \langle \{f, y\} \rangle, |\langle y \rangle| = n\} \,. \tag{25}$$

From Theorem 4.9 follows that $C_1(n) \leq (n-1)^2$. On the other hand, it is well known that $\mathrm{rt}(A) = (n-1)^2$ for the synchronizing $n$-states Černý automaton $A = (S, \{f, y\})$ [10]. Hence, we have $C_1(n) \geq \mathrm{rt}(M_A) \geq \mathrm{rt}(A) = (n-1)^2$. From this and the inequality $C_1(n) \leq (n-1)^2$ we obtain the following statement.

**Corollary 4.10.** The following equality holds $C_1(n) = (n-1)^2$, for $n \geq 2$.

Now we consider the general case when the cyclic group $\langle y \rangle$, where $y$ satisfies Eq. (10), is a subgroup of the regular group $G$ in a Černý type monoid $M = \langle \{f\} \cup G \rangle$. Let us put $H = \langle y \rangle$, $q = |H|$, $r = |G:H|$. Consider the equivalence $\varepsilon$, whose classes $S_i$, $1 \leq i \leq r$, was defined by the Eq. (11). Note that the partition $S/\varepsilon = \{S_1, \cdots, S_r\}$ is uniform, since $|S_i| = |H| = q$, $1 \leq i \leq r$. Hence, the following equality holds:

$$q \cdot r = n \,. \tag{26}$$

According lemma 4.1 the equivalence $\varepsilon$ is a t-invariant congruence of a monoid $M = \langle \{f\} \cup G \rangle$. From Eq. (12) we see that $f$ acts identically on the partition $S/\varepsilon$. Then the map $w|_\varepsilon : S/\varepsilon \to S/\varepsilon$ will be a permutation on $S/\varepsilon$ for each $w \in M$:

$$w|_\varepsilon : S/\varepsilon \leftrightarrow S/\varepsilon \,. \tag{27}$$

in this case, the transformation factor-monoid $M|_\varepsilon$ equals to the group $G|_\varepsilon$. Moreover, the group $G|_\varepsilon$ is transitive, because $G$ is transitive. Hence, for any class $S_i$, $2 \leq i \leq r$, there is a permutation $h_i \in G$, such that:

$$(S_i)h_i = S_1 \,. \tag{28}$$

**Theorem 4.11.** Let $M = \langle \{f\} \cup G \rangle$ be a Černý type monoid with $n$ states $|S| = n$, then $\mathrm{rk}(M) = r = |G : H|$ and $\mathrm{rt}(M) \leq (n - r)^2/r + C_r^2$, where $C_r^2 = r(r - 1)/2$ is the binomial coefficient.

**Proof.** Let $X$ be a set of generators of $M = \langle \{f\} \cup G \rangle$, for which $\mathrm{rt}_X(M) = \mathrm{rt}(M)$. We divide $X$ into two parts $X = Y \cup Z$, where $Y$ are permutations, which generates the group $G = \langle Y \rangle$ and $Z$ are singulars. Consider the class $T = S_1 = (e_f)H$ of the congruence $\varepsilon$, then from Eq. (12) follows that $(T)f \subset T$. Moreover, we have $(T)y = (e_f)Hy = (e_f)H = T$. Hence, the class $T$ is invariant for the submonoid $M_1 = \langle \{f, y\} \rangle$, so $M_1$ acts on the class. From Theorem 4.5 follows that $M_1|_T = \langle \{f|_T, y|_T\} \rangle$ will be the Černý type synchronizing monoid with $q = |T|$ states. Thus, from Theorem 4.9 we conclude that there is a map $w \in M_1$ such that:

$$|(S_1)w| = |(T)w| = 1, \quad l_X(w) \leq (q - 1)^2. \tag{29}$$

From this, taking into account Eq. (26), we obtain the estimation:

$$l_X(w) \leq (q - 1)^2 = \left(\frac{n}{r} - 1\right)^2 = \frac{(n - r)^2}{r^2}. \tag{30}$$

According Eq. (28) we define the distance $d(S_i)$, $2 \leq i \leq r$, from $S_i$ to $S_1$ by the formula $d(S_i) = \min\{l_Y(h_i) : (S_i)h_i = S_1\}$. Then we put the classes in order by the distance $0 < d(S_2) \leq \cdots \leq d(S_r)$, and suppose that $l_Y(h_i) = d(S_i)$, $2 \leq i \leq r$. Thus, we obtain the following conditions:

$$(S_i)h_i = S_1, \quad l_Y(h_i) \leq i - 1, \quad 2 \leq i \leq r. \tag{31}$$

Further, we construct the maps $w_i$, $1 \leq i \leq r$, step by step. In the beginning, we put $w_1 = w$ and $T_1 = S_1$, then from Formula (29) we have $|(S_1)w_1| = |(T_1)w_1| = 1$. Note that the rest classes $\{S_2, \cdots, S_r\}$ are not merged, since the map $w_1 \in \langle \{f, y\} \rangle$ permutes them. On the step $j$, $2 \leq j \leq r$, when $j - 1$ classes are merged, let $S_{i_j}$ be the class with the minimal number for which $S_{i_j} \subseteq \mathrm{im}(w_{j-1})$, then $2 \leq i_j \leq j$. According Formula (27) there is the unique class $T_j$ such that $(T_j)w_{j-1} = S_{i_j}$. Then we put $w_j = w_{j-1}h_{i_j}w$, and obtain the equality:

$$|(T_j)w_j| = |(T_j)w_{j-1}h_{i_j}w| = \left|(S_{i_j})h_{i_j}w\right| = |(S_1)w| = 1. \tag{32}$$

As a result, we obtain the map $w_r = wh_{i_2}w \cdots wh_{i_r}w$, which transfers each class $T_j$, $1 \le j \le r$ into one state on the strength of Eq. (32). Hence, $\text{rk}(M) \le \text{rk}(w_r) \le r$. On the other hand, from Corollary 4.2 we have $\text{rk}(M) \ge r$. From this, we conclude that $\text{rk}(w_r) = \text{rk}(M) = r = |G:H|$.

By construction we have $1 \le i_j \le j$, $2 \le j \le r$, then from Properties (6), (30) and (31) we obtain the following upper bound:

$$\text{rt}(M) = \text{rt}_X(M) \le l_X(w_r) \le r \cdot l_X(w) + \sum_{j=2}^{r} l_Y(h_{i_j}) \le \frac{(n-r)^2}{r} + C_r^2.$$

Thus, the theorem is proved. □

Let us compare the upper bound from this theorem with the number $(n-r)^2$, then taking into account (26) we have:

$$(n-r)^2 - \frac{(n-r)^2}{r} - C_r^2 = (r-1)\frac{2(n-r)^2 - r^2}{2r} =$$

$$= (r-1)\frac{(n-r)^2 + n(n-2r)}{2r} = (r-1)\frac{(n-r)^2 + nr(q-2)}{2r} \ge 0,$$

since $r \ge 1$ and $q \ge 2$. From this, we obtain the statement.

**Corollary 4.12.** A Černý type monoid $M = \langle\{f\} \cup G\rangle$ satisfies the rank conjecture (Conjecture 3.2).

Now we turn to obtaining the lower bound for the rank threshold of a Černý type monoid. Let $Q = \{0,1,\cdots q-1\}$, $R = \{0,1,\cdots r-1\}$ and $n = q \cdot r$, then consider the automaton $A = (S, \{f, y, z\})$, where $S = Q \times R$ is a set of states with $n$ elements. We define the simple idempotent $f$ as follows $(q-1,0)f = (0,0)$ and stable on all other states. The permutations $y$ and z are defined in the following way for each $(i,j) \in S$:

$$(i,j)y = ((i+1) \bmod q, j), \quad (i,j)z = (i, (j+1) \bmod r).$$

Thus, the group $G = \langle\{y,z\}\rangle$ is commutative and transitive, since the following properties hold for all $(i,j) \in S$, $0 \le i \le q-1$ and $0 \le j \le r-1$:

$$(i,j)yz = ((i+1) \bmod q, (j+1) \bmod r) = (i,j)zy, \quad (0,0)y^i z^j = (i,j).$$

Hence, the group $G$ is regular. Indeed, if $(s)g = (s)h$ for some $s \in S$, then for any $t \in S$ there is a permutation $g_t \in G$ such that $(s)g_t = t$. Therefore, we have $(t)g =$ $(s)g_tg = (s)gg_t = (s)hg_t = (t)h$, for each $t \in S$, hence $g = h$. Thus, the monoid $M_A = \langle \{f\} \cup G \rangle$ is Černý type.

Consider the subset of states $T = Q \times \{0\}$, then $(T)f \subset T$ and $(T)y = T$. So, the sub-automaton $A_1 = (T, \{f|_T, y|_T\})$ with $q$ states will be isomorphic to the Černý automaton [10]. Ley us put $h = y^{q-1}$, then we have:

$$|(T)w| = 1, \text{ where } = (fh)^{q-2}f. \tag{33}$$

Let $H = \langle y \rangle$ be a cyclic subgroup of $G$ and $H_j = Hz^j$, $0 \leq j \leq r-1$, be the cosets of the group $G$ by the subgroup $H$. Consider the equivalence $\varepsilon$, whose classes $S_j$ are defined by formula $S_j = (q-1, 0)H_j = (q-1, j)H$, $0 \leq j \leq r-1$. Then by Lemma 4.1 $\varepsilon$ is a t-invariant congruence of the transition monoid $M_A$. So according Corollary 4.2 we have $\text{rk}(M_A) \geq |G:H| = r$. On the other hand, the following properties follows from Eq. (33), since $S_0 = T$:

$$|(S_0)w| = 1, \ (S_j)f = (S_j)y = S_j, \ (S_j)z^{r-j} = S_0, \ 1 \leq j \leq r-1. \tag{34}$$

From this, we conclude that the map $w_r = w(zw)^{r-1}$ merges each class $S_j$ into one state $|(S_j)w_r| = |(S_j)z^{r-j}w| = 1$, $0 \leq j \leq r-1$. Thus, we have $\text{rk}(M_A) = \text{rk}(w_r) = r$. Let us put $X = \{f, y, z\}$ then we obtain by construction:

$$\text{rt}(A) \leq l_X(w_r) = r(q-1)^2 + r - 1 = \frac{(n-r)^2}{r} + r - 1. \tag{35}$$

Now let $v \in M_A$ be a map of the rank $r$ and $l_X(v) = \text{rt}(A)$. Consider its shortest factorization $v = w_0 g_1 \cdots w_{k-1} g_k w_k$ in respect to $X$, where $w_i \in \langle \{f, y\} \rangle$, $0 \leq i \leq k$, and $g_i \in \langle z \rangle \setminus \{1_S\}$, $1 \leq i \leq k$. In this case, we have the equality:

$$\text{rt}(A) = l_X(v) = \sum_{i=0}^{k} l_X(w_i) + \sum_{i=1}^{k} l_X(g_i). \tag{36}$$

We divide the maps $w_i$, $0 \leq i \leq k$, by collections $W_j$, $0 \leq j \leq r-1$, in the following way. We put $w_0 \in W_0$ and $w_i \in W_j$, if $g_1 \cdot \cdots \cdot g_i = z^{r-j}$, $1 \leq i \leq k$. Then all collections are nonempty, since each class $S_j$, $1 \leq j \leq r-1$, has to reach the class $S_0$ and collections are not intersect. Hence, by the box principle we have:

$$k \geq r - 1. \tag{37}$$

Let $v_j$ be the product of all maps from $W_j$, $0 \leq j \leq r-1$, taken in the natural order, then from Eq. (34) follows that $|(S_j) z^{r-j} v_j| = |(S_0) v_j| = 1$, since $|(S_j) v| = 1$, $0 \leq j \leq r-1$. Hence, all maps $v_j|_T$ are constants in the Černý automaton $A_1$, then we conclude from [10] that $l_X(v_j) \geq (q-1)^2$, $0 \leq j \leq r-1$. Therefore, taken into account Properties (6), (36) and (37), we obtain the lower bound for the rank threshold of the automaton $A$:

$$\text{rt}(A) = l_X(v) = \sum_{i=0}^{k} l_X(w_i) + \sum_{i=1}^{k} l_X(g_i) \geq \sum_{j=0}^{r-1} l_X(v_j) + k \geq r(q-1)^2 + r - 1.$$

From this and Property (35) we deduce the rank threshold of the automaton $A$:

$$\text{rt}_X(M_A) = \text{rt}(A) = r(q-1)^2 + r - 1 = \frac{(n-r)^2}{r} + r - 1. \quad (38)$$

We generalize the Černý function, which was defined by Formula (25), in the following way:

$$C_r(n) = \max_M \{\text{rt}(M) : M = \langle \{f\} \cup G \rangle, \ |G| = n, \ \text{rk}(M) = r\}.$$

From Eq. (38) and Theorem 4.11 follows the lower and upper bounds for this function:

$$\frac{(n-r)^2}{r} + r - 1 \leq C_r(n) \leq \frac{(n-r)^2}{r} + C_r^2.$$

Note that the difference between the upper and the lower bounds equals to the binomial coefficient $C_{r-1}^2 = (r-1)(r-2)/2$, which does not depend on $n$.

**5. Conclusion**

This paper confirms the thesis stated in [5] that monoidal automata allow a deeper use of the theories semigroups and groups in solving the Černý problem. Further research may involve generalizing these results and in particular Theorem 4.11 to monoids with a transitive permutation group.

**Vitae**

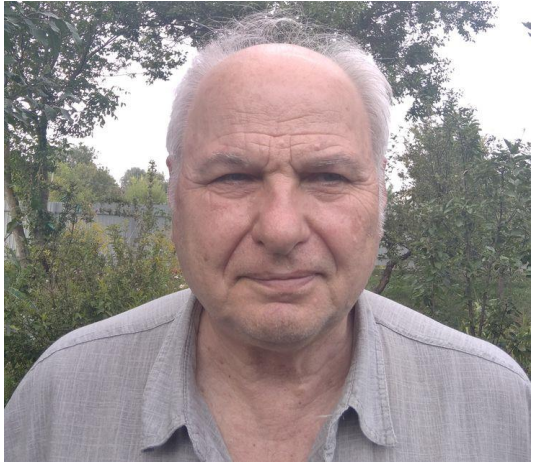

Rystsov Igor Doctor of Science in mathematics, Professor (Associate) at National Technical University of Ukraine (Kiev Polytechnic Institute).

Skills:

Automata Theory,·Discrete Mathematics, Algebra.